\def\aa{A\&A}               %Astronomy & Astrophysics%
\def\aas{A\&AS}             %A & A Supplements%
\def\anj{AJ}                %Astronomical Journal%
\def\apj{ApJ}               %Astrophysical Journal%
\def\apjs{ApJS}             %Astrophysical Journal Supplements
\def\mn{MNRAS}              %Monthly Notices of the Royal...%
\def\pasp{PASP}             %Publ. of the Astr. Soc. of the Pacific%
\def\newar{NewAR}           %New Astronomy Reviews

\documentclass{aa}
\begin{document}

%\thesaurus{3(11.01.2; 11.09.1 J1835+620; 11.10.1; 13.18.1)} 

\title{A new sample of large angular size radio galaxies}
\subtitle{I. The radio data}

%\author{L. Lara et al.}

\author{L. Lara\inst{1} \and 
W.D. Cotton\inst{2} \and
L. Feretti\inst{3} \and
G. Giovannini\inst{3,4} \and
J.M. Marcaide\inst{5} \and
I. M\'arquez\inst{1} \and
T. Venturi\inst{3}}

\offprints{L. Lara,
\email{lucas@iaa.csic.es}}

\institute{Instituto de Astrof\'{\i}sica de Andaluc\'{\i}a (CSIC),
Apdo. 3004, 18080 Granada, Spain
\and 
National Radio Astronomy Observatory, 520 Edgemont Road, Charlottesville, 
VA 22903-2475, USA
\and
Istituto di Radioastronomia (CNR), via P. Gobetti 101, 40129 Bologna, Italy
\and
Dipartimento di Fisica, Universit\'a di Bologna, via B. Pichat 6/2,
40127 Bologna, Italy
\and
Departamento de Astronom\'{\i}a, Universitat de Val\`encia, 46100 Burjassot, 
Spain
} 
\date{Received / Accepted}

\authorrunning{Lara et al.}
\titlerunning{A new sample of large angular size radio galaxies. I}

\abstract{
We present a new sample of 84 large angular size radio galaxies
selected from the NRAO VLA Sky Survey. Radio sources with declination
above $+60^{\circ}$, total flux density greater than 100 mJy at 1.4
GHz and angular size larger than 4$\arcmin$ have been selected and
observed with the VLA at 1.4 and 4.9 GHz. The radio observations
attempt to confirm the large angular size sources and to isolate the core 
emission for optical identification. In this paper, the first of a
series of three, we present radio maps of 79 sources from the sample and
discuss the effects of the selection criteria in the final sample. 
37 radio galaxies belong to the class of giants, of which 
22 are reported in this paper for the first time.
}

\maketitle

%
%--------------------------------------------------------------------------

\section{Introduction}

The NRAO\footnote{National Radio Astronomy Observatory}
VLA\footnote{Very Large Array, operated by the NRAO} Sky Survey (NVSS;
Condon et al. \cite{nvss}) provides radio maps of the sky north of
$-40^{\circ}$ declination, at a frequency of 1.4 GHz, in total and
polarized intensity, with an angular resolution of 45$\arcsec$ and 1
r.m.s. = 0.45 mJy/beam.  Due to its high sensitivity and resolution
compared with other all-sky surveys, the NVSS is a unique tool for the
definition of complete samples of extended objects, not known before
because of their low flux density and/or because of confusion effects
present in previous surveys.

In 1995 we undertook a project for the definition and study of a
sample of large angular size radio galaxies selected from the NVSS
with the following aims:

{\em i)} To construct a sample of radio galaxies with their jets
oriented near the plane of the sky and study the parsec scale
properties of these jets. Most samples selected for Very Long Baseline
Interferometry (VLBI) observations are usually defined on the basis of
a flux density cutoff at high frequencies (eg. Pearson \& Readhead 
\cite{pearson1}, Polatidis et al. \cite{polatidis}, Taylor et al. 
\cite{taylor}, Kellermann et al. \cite{kellermann}, 
Fomalont et al. \cite{fomalont}), although
it is known that the combined effect of such cutoff and the Doppler
boosting of the radio emission introduces a bias in those samples
towards small orientation angles with respect to the observer's line
of sight.  As a consequence, only a few sources with symmetric
structure on the parsec scale are currently known, despite the
importance of these sources to understand the physical properties of
parsec scale jets. For the selection of our sample we did not impose
any core flux density limitation, and considered that simply due to
projection effects, a significant number of large angular size radio
sources would be oriented close to the plane of the sky and should
show symmetric parsec scale jets.

{\em ii)} To study the properties of giant radio galaxies (GRGs;
defined as those with a projected linear size $\ge$ 1 Mpc\footnote{We
assume that H$_{0}=50$~km~s$^{-1}$~Mpc$^{-1}$and q$_{0}$=0.5
throughout this paper.}). This kind of source is difficult to detect:
in Fanaroff-Riley type II (FR II; Fanaroff \& Riley
\cite{fanaroff}) giant sources, the association between the two
hotspots is often not obvious to establish; on the other hand, the low
brightness extended lobes in Fanaroff-Riley type I (FR I) sources can
only be detected with deep radio observations and therefore, such
sources are missed in most radio surveys.  Ishwara-Chandra \&
Saikia (\cite{ishwara}) collected a sample of 53 GRGs from
inhomogeneous literature data, but only recently, the NVSS and the
Westerbork Northern Sky Survey at 325 MHz (WENSS; Rengelink et
al. \cite{wenss}) provide the sufficient sensitivity to detect GRGs in
a systematic way.  Schoenmakers (\cite{arnotesis}) presented a sample
of 47 GRGs from the WENSS survey. A new sample of large angular size
radio galaxies will contain an appreciable number of intrinsically
large radio galaxies and will allow us to increment the number of
known GRGs. A large sample with homogeneous selection criteria is
necessary to study the poorly known properties of GRGs, and how their
jets interact with the external medium at large distances from the
parent galaxy (Ishwara-Chandra \& Saikia \cite{ishwara}; Schoenmakers
et al. \cite{arno2}; Lara et al. \cite{lara3}).

{\em iii)} To investigate the evolution of radio galaxies. The
selection of a sample of large angular size radio galaxies and the
comparison with other samples will give new information on the time
evolution of radio sources, since a larger source size implies, in
principle, a higher probability of selecting older sources than in
other samples (Ishwara-Chandra \& Saikia \cite{ishwara}; 
Schoenmakers et al. \cite{arno2}).

We present here (paper I) the sample definition and members, VLA radio
maps of 79 selected objects and the main physical parameters of the
sources, and we discuss the effects of the selection criteria on the
sample.  Papers II and III (in preparation) will present optical data
(images and spectroscopy) of the associated galaxies and an analysis
of the sample properties with statistical considerations,
respectively. 

\section{Sample pre-selection}

We considered all the NVSS maps above $+60^{\circ}$ of declination
(covering an area of $\pi$ steradians) and proceeded to a careful
visual inspection of the contour map plots (at a 2$\sigma$ level).
Each $4^{\circ}\times4^{\circ}$ NVSS map was divided in 16
$1^{\circ}\times1^{\circ}$ maps to facilitate the search for
candidates.  We then pre-selected those map features apparently
related to a single source, with a total flux density $\ge 100$ mJy
and an angular extension larger than 4$\arcmin$.

Flux density measurements on the NVSS maps were done using the task
TVSTAT in AIPS\footnote{Astronomical Image Processing System,
developed and maintained by the NRAO}, defining a polygonal area
embracing all the source emission. For the computation of the angular
extension we considered the maximum distance between contours at
3$\sigma$ level when the source was extended and diffuse with no
evident sub-structure, or the distance between peaks of brightness
when there were unresolved features (hotspots) at the source
extremes. We took into account, when possible, curvatures in the radio
structure of the sources so that a measured angular size corresponds to  
the length along the ``spine'' of the radio source. 
A total of 122 sources were pre-selected for subsequent
confirmation through higher resolution radio observations.

We note that due to the existence of ``holes'' in the NVSS
maps at the time of our search of candidates, a few sources fulfilling
the requirements of the sample may have been skipped in the selection
process and missed from the final sample.  In fact, the known giant
radio galaxies 4C+73.08 (J0949+732) and 4C+74.26 (J2042+751) were not 
selected for the
pre-sample at this first stage, but were added later to the final
sample. We note however that possible missing sources will not affect the
statistical considerations derived from the sample.

\section{Observations and final sample selection}

All the 122 pre-selected sources were
observed with the VLA between 1995 and 1998 (see Table~\ref{obs} for
details). At least one snapshot at 1.4 and 4.9 GHz was obtained for
each source.  When possible, we obtained for the same source snapshots
in B- and C-configurations at both frequencies to increase the image
quality by combining the two different configuration data.  The radio
sources 3C\,286 and/or 3C\,48 served as primary flux density
calibrators.  The interferometric phases were calibrated using nearby
radio sources selected from the VLA calibrator manual (Perley \&
Taylor \cite{perley2}). The processes of self-calibration and imaging
of the data were carried out with the NRAO AIPS package, following
standard procedures. Due to the large size of the sources, all the
maps made at 4.9 GHz and, when necessary, also at 1.4 GHz, were
corrected for primary beam attenuation.

\begin{figure}
\vspace{12cm}
\includegraphics{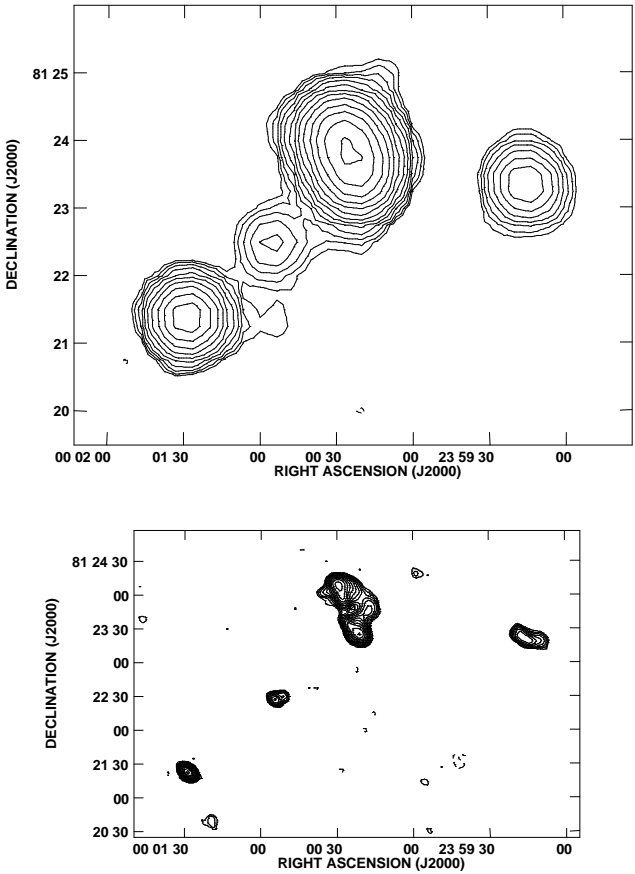}
%\rule{0.4pt}{4cm}% line thickness, height of picture
\caption{Illustration of the necessity of high resolution observations of NVSS
sources in order to discard some objects resulting from superposition
of several unrelated sources. {\bf Top:} an image of a presumed large
size double radio galaxy selected from the NVSS. {\bf Bottom:} the
corresponding image obtained from VLA B-configuration observations at
1.4 GHz.}
\label{fig_cheat}
\end{figure}

The aims of the observations were to confirm large angular size radio
galaxies, reject those objects which were in fact the result of the
superposition of two or three adjacent sources (see
Fig.~\ref{fig_cheat}) and isolate the core emission to obtain
accurate positions for subsequent optical identification and redshift
determination. 

\begin{figure*}
\vspace{23cm}
\includegraphics{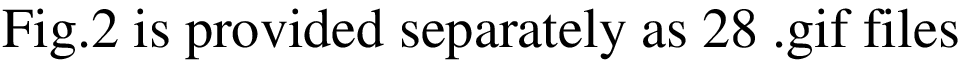}
%\rule{0.4pt}{4cm}% line thickness, height of picture
\caption{VLA maps of 79 radio galaxies from the sample.  A label ``L'' or
``C'' by the source name indicates the frequency of observation,
namely 1.4 or 4.9 GHz, respectively. Dotted lines help to identify
equivalent regions in different frequency maps. The core position is
marked with an arrow on each map. The angular scale of the 1.4 GHz
maps has been kept fixed, except for objects too extended to properly
fit in a single page. In those cases, the map size is reduced by a
factor of 2 or 3 (marked with ``1/2'' or ``1/3'', respectively). Map 
parameters are displayed 
in Table~\ref{parameters}. Contours are separated by factors of $\sqrt{2}$}
\label{mapas}
\end{figure*}

82 of the 122 pre-selected radio sources were identified beyond doubt
with single radio galaxies. 13 of those 82 sources turned out to have
a size slightly smaller than 4$\arcmin$ after accurate measurements
from higher resolution observations.  Another source has a total flux
density at 1.4 GHz below 100 mJy after the subtraction of the flux
density from a closeby unrelated source.  However, we have kept them
all in the final sample since this fact will not influence our
considerations. Although the integration time (10 minutes on the
average) was not too long on each single source, thanks to the use of
combined arrays and to the good uv-coverage of the VLA observing with
the snapshot technique, we obtained the high quality images shown in
Fig.~\ref{mapas}. We present only total intensity data.

In Table~\ref{sample} we report the most relevant physical parameters
of the 84 radio galaxies in the final sample (82 selected objects plus
the two forementioned giants 4C+73.08 and 4C+74.26). Coordinates are
derived from Gaussian fits of the core component at our highest
angular resolution observations (commonly 4.9 GHz,
B-configuration). Total flux densities at 1.4 GHz (S$_t^{1.4}$) are
from the NVSS, except when a higher resolution image is necessary to
isolate and subtract discrete unrelated sources. Angular sizes are 
measured along the spine of the sources using our 
VLA maps at 1.4 GHz, or the NVSS maps when there is extended emission
resolved out in our observations.
The core flux
density at 1.4 GHz (S$_c^{1.4}$) and 4.9 GHz (S$_c^{4.9}$) has been
estimated fitting a Gaussian component to the core using AIPS task
JMFIT. The peak flux density of the Gaussian was taken as the core
flux density. The core peak flux densities of those sources whose cores
have very different peak and integrated fluxes (peak to integrated
ratio $<$ 0.8) have been marked with an ``e'' in
Table~\ref{sample}. Known redshifts have been compiled from the NASA
Extragalactic Database (NED) and from Schoenmakers
(\cite{arnotesis}). 44 new redshifts have been derived from
our optical observations (Paper II). At present, the distance
information is available for 71 of the sources (6 of which have
uncertain redshift determination).  When a redshift measure is
available, we give also physical results as the projected linear size,
the total radio power at 1.4 GHz (P$_t^{1.4}$) and the core radio
power at 1.4 GHz (P$_c^{1.4}$) and 4.9 GHz (P$_c^{4.9}$).

\begin{table}[]
\caption[]{VLA observations}
\label{obs}
\begin{tabular}{lrrrc}
\hline
Array  & $\nu$  & $\Delta\nu$ & Date & Code$^a$ \\
       & (MHz)  & (MHz)       &      &      \\
\hline
 B     &  1465  &  50 & 19 Nov 95    & {\em a} \\
 C     &  1465  &  50 & 19 Feb 96    & {\em b} \\
 C     &  4885  & 100 & 19 Feb 96    & {\em c} \\
 B     &  1418  &  50 & 18-21 Feb 97 & {\em d} \\
 B     &  4885  & 100 & 18-21 Feb 97 & {\em e} \\
 B     &  1452  &  50 & 25 May 97    & {\em f} \\
 B     &  4885  & 100 & 25 May 97    & {\em g} \\
 C     &  1465  &  50 & 27 Jul 97    & {\em h} \\
 C     &  4885  & 100 & 27 Jul 97    & {\em i} \\
 B     &  1465  &  50 &  7 Oct 98    & {\em j} \\
 B     &  4885  & 100 &  7 Oct 98    & {\em k} \\
 C     &  1465  &  50 & 28 Dec 98    & {\em l} \\ 
 C     &  4885  & 100 & 28 Dec 98    & {\em m} \\
\hline 
\end{tabular}
\parbox{7cm}{
$^a$ Letter code used in Tab.~\ref{parameters} to show the observations 
made on each source.}
\end{table}

\begin{table*}[t]
\caption[]{Radio sample of large angular size radio galaxies from the NVSS survey}
\label{sample}
%\begin{footnotesize}
\begin{scriptsize}
\begin{tabular}{lccrr r@{.}l r@{.}l lrrrrl}
~~~Name & R.A.(J2000.0)&Dec.(J2000.0)  & S$_{t}^{1.4}$ & Size & \multicolumn{2}{c}{S$_{c}^{1.4}$} & \multicolumn{2}{c}{S$_{c}^{4.9}$} & ~~z& Size & P$_{t}^{1.4}$ & P$_{c}^{1.4}$ &  P$_{c}^{4.9}$ & Type$^{\ast}$ \\ 
          &($^h~~^m~~^s$)&($^{\circ}$~~$\arcmin$~~$\arcsec$)& (mJy)&  ($\arcmin$) & \multicolumn{2}{c}{(mJy)} & \multicolumn{2}{c}{(mJy)} & & (Kpc)& W/Hz &  W/Hz & W/Hz & \\
\hline	 
\object{J0109+731} & 01 09 44.265 & 73 11 57.17 & 3030 &  4.0 & 22&0 & 10&0 & 0.181$^a$ &   943 & 26.67 & 24.53 & 24.18 & II\\
\object{J0153+712} & 01 53 25.786 & 71 15 06.53 & 1200 &  7.0 &512&3 &292&2 & 0.022$^a$ &   259 & 24.40 & 24.03 & 23.79 & I \\
\object{J0317+769} & 03 17 54.061 & 76 58 37.82 &  196 &  3.4 & 25&3e& 11&9 & 0.094     &   476 & 24.89 & 24.00 & 23.67 & I \\
\object{J0318+684} & 03 18 19.026 & 68 29 32.08 &  823 & 15.1 & 22&0 & 44&0 & 0.090$^b$ &  2040 & 25.48 & 23.90 & 24.20 & II\\
\object{J0342+636} & 03 42 10.148 & 63 39 33.73 &  284 &  4.1 & 10&4 & 11&3 & 0.128     &   741 & 25.33 & 23.89 & 23.93 & II\\
{\bf \object{J0430+773}} & 04 30 49.490 & 77 22 58.44 &  150 &  4.9 &  3&3 &  5&6 & 0.215     &  1306 & 25.52 & 23.86 & 24.09 & II\\
\object{J0455+603} & 04 55 47.039 & 60 23 36.17 &  109 &  4.0 & 21&5 &  7&3 & ---       &   --- &   --- &   --- &  ---  & I \\
\object{J0502+670} & 05 02 54.732 & 67 02 30.15 &  104 &  4.8 &  6&9 &  2&3 & 0.085     &   617 & 24.53 & 23.35 & 22.87 & I \\
{\bf \object{J0508+609}} & 05 08 27.258 & 60 56 27.48 &  221 & 10.4 & 71&0 & 74&8 & 0.071     &  1142 & 24.69 & 24.20 & 24.22 & I \\
\object{J0519+702} & 05 19 17.132 & 70 13 48.68 &  254 &  4.8 & 18&9e& 11&3 & 0.144     &   952 & 25.38 & 24.26 & 24.03 & I \\
\object{J0525+718} & 05 25 27.094 & 71 52 39.25 &   69 &  4.6 &  4&7e&  1&5 & 0.150     &   942 & 24.85 & 23.68 & 23.19 & I \\
\object{J0531+677} & 05 31 25.925 & 67 43 50.23 &  210 &  9.6 &106&6e& 10&6e& 0.017     &   276 & 23.42 & 23.13 & 22.12 & I \\
\object{J0546+633} & 05 46 24.622 & 63 21 32.50 &  392 & 11.0 & 29&0 & 13&5 & 0.049$^a$ &   865 & 24.62 & 23.49 & 23.15 & I \\
\object{J0559+607} & 05 59 38.690 & 60 44 00.96 &  115 &  7.1 &  2&5 &  2&2 & 0.042     &   484 & 23.95 & 22.29 & 22.23 & I \\
{\bf \object{J0607+612}} & 06 07 34.919 & 61 14 43.52 &  358 &  5.3 &  4&5 &  4&1 & 0.227     &  1465 & 25.94 & 24.04 & 24.00 & ?\\
\object{J0624+630} & 06 24 29.063 & 63 04 02.50 &  181 &  5.4 &  3&5 &  4&4 & 0.085     &   694 & 24.77 & 23.05 & 23.15 & I \\
\object{J0633+721} & 06 33 40.842 & 72 09 24.92 &  308 &  5.0 & 14&0 & 12&0 & 0.090$^b$ &   675 & 25.05 & 23.71 & 23.64 & ?\\
\object{J0654+733} & 06 54 26.525 & 73 19 50.36 &  868 & 12.2 &  3&2 &  4&1 & 0.115$^b$ &  2015 & 25.71 & 23.28 & 23.39 & II\\
{\bf \object{J0750+656}} & 07 50 34.425 & 65 41 25.50 &  120 &  3.7 & 26&8 & 29&4 & 0.747     &  1798 & 26.57 & 25.92 & 25.96 & II\\
\object{J0757+826} & 07 57 35.172 & 82 39 40.86 &  153 &  4.0 & 33&7 & 17&1e& 0.087?    &   524 & 24.71 & 24.06 & 23.76 & I \\
{\bf \object{J0803+669}} & 08 03 45.829 & 66 56 11.39 &  182 &  4.7 &  5&2 &  3&8 & 0.247?    &  1374 & 25.73 & 24.18 & 24.04 & II\\
\object{J0807+740} & 08 07 10.070 & 74 00 41.58 &  152 &  9.1 & 14&0 & 12&4 & 0.120$^b$ &  1566 & 25.00 & 23.96 & 23.91 & ? \\
\object{J0819+756} & 08 19 50.504 & 75 38 39.53 &  616 &  7.8 & 30&7 & 56&6 & 0.232$^b$ &  2190 & 26.20 & 24.90 & 25.16 & II\\
\object{J0825+693} & 08 25 59.770 & 69 20 38.59 &  141 &  7.2 &  7&1 &  7&1 & 0.538$^a$ &  3162 & 26.33 & 25.03 & 25.03 & II\\
\object{J0828+632} & 08 28 56.363 & 63 13 45.05 &  182 &  4.0 &  6&3 &  3&7 & ---       &   --- &   --- &   --- &  ---  & ?\\
{\bf \object{J0856+663}} & 08 56 16.260 & 66 21 26.50 &  236 &  3.8 &  5&3 & 10&3 & 0.489     &  1607 & 26.47 & 24.82 & 25.11 & II\\
{\bf \object{J0926+653}} & 09 26 00.822 & 65 19 22.88 &  101 &  5.3 &  1&0e&  2&2e& 0.140     &  1028 & 24.96 & 22.95 & 23.30 & I \\
{\bf \object{J0926+610}} & 09 26 53.408 & 61 00 24.87 &  300 &  3.7 &  6&8 &  7&2 & 0.243     &  1070 & 25.93 & 24.28 & 24.31 & II\\
\object{J0939+740} & 09 39 46.833 & 74 05 30.78 &  100 &  8.0 &  3&9 &  4&5 & 0.122$^b$ &  1387 & 24.83 & 23.42 & 23.48 & I \\
\object{J0949+732}$^c$ & 09 49 46.157 & 73 14 23.82 & 2537 & 15.0 &15&6 &20&0 &0.058$^a$ & 1375 & 25.58 & 23.37 & 23.47 & II\\
{\bf \object{J1015+683}} & 10 15 21.620 & 68 23 58.24 &  390 &4.0&\multicolumn{2}{c}{---}&5&6& 0.199 &1010&25.86 & ---& 24.02 & ?\\
\object{J1036+677} & 10 36 41.237 & 67 47 53.44 &  231 &  4.0 &  2&6 &  3&8 &  ---      &   --- &   --- &   --- &   --- & II\\
\object{J1124+749} & 11 24 47.045 & 74 55 45.31 &  125 &  5.6 & 38&4 & 23&8 & 0.052     &   465 & 24.17 & 23.66 & 23.45 & I \\
\object{J1137+613} & 11 37 21.289 & 61 20 01.88 & 1160 &  3.4 & 13&1e& 12&5 & 0.111     &   549 & 25.81 & 23.87 & 23.85 & II \\
\object{J1211+743} & 12 11 58.710 & 74 19 04.12 &  628 &  7.5 &  9&7 & 11&3 & 0.107     &  1171 & 25.51 & 23.70 & 23.77 & ? \\
{\bf \object{J1216+674}} & 12 16 37.239 & 67 24 41.97 &  187 &  5.8 &  2&0 &  3&2 & 0.362     &  2126 & 26.09 & 24.12 & 24.32 & II\\
\object{J1220+636} & 12 20 36.477 & 63 41 43.82 &  261 &  5.2 &  4&2 &  6&4 &  ---      &   --- &   --- &   --- &   --- & II\\
\object{J1247+673} & 12 47 33.319 & 67 23 16.34 &  388 & 12.4 &262&4 &179&0 & 0.107$^a$ &  1938 & 25.30 & 25.13 & 24.97 & II\\
{\bf \object{J1251+756}} & 12 51 05.977 & 75 37 38.94 &  100 &  4.0 &  2&9 &  4&4 & 0.197     &  1002 & 25.26 & 23.72 & 23.90 & II\\
\object{J1251+787} & 12 51 23.839 & 78 42 36.29 &  166 & 19.5 & 13&2 & 15&5 &  ---      &   --- &   --- &   --- &   --- & I \\
\object{J1313+696} & 13 13 58.878 & 69 37 18.74 & 1384 &  6.9 & 10&2e&  3&8 & 0.106$^a$ &  1062 & 25.85 & 23.71 & 23.29 & II\\
\object{J1410+633} & 14 10 30.609 & 63 19 00.55 &  249 &  3.4 &  3&1e&  1&1e& 0.158     &   724 & 25.46 & 23.55 & 23.10 & II\\
\hline 
\end{tabular}
\parbox{16cm}{
$^{\ast}$ I and II stand for FR I and FR II type radio galaxies, respectively\\
$^a$ Redshift taken from the NASA Extragalactic Database\\
$^b$ Redshift from Schoenmakers (\cite{arnotesis})\\
$^c$ Coordinates taken from Saripalli et al. (\cite{saripalli})\\
Label ``e'' by the core flux density means extended core (see text).\\
New giant radio galaxies are written in boldface. 
}
%\end{footnotesize}
\end{scriptsize}
\end{table*}

\begin{table*}[t]
\addtocounter{table}{-1}
\caption[]{({\em cont.}) Radio sample of large angular size radio galaxies from the NVSS survey}
\begin{scriptsize}
\begin{tabular}{lccrrr@{.}l r@{.}l lrrrrl}
~~~Name      & R.A.(J2000.0)&Dec.(J2000.0)  & S$_{t}^{1.4}$ & Size & \multicolumn{2}{c}{S$_{c}^{1.4}$} & \multicolumn{2}{c}{S$_{c}^{4.9}$} & ~~z& Size & P$_{t}^{1.4}$ & P$_{c}^{1.4}$ &  P$_{c}^{4.9}$ & Type$^{\ast}$ \\ 
          &($^h~~^m~~^s$)&($^{\circ}$~~$\arcmin$~~$\arcsec$)& (mJy)&  ($\arcmin$) & \multicolumn{2}{c}{(mJy)} & \multicolumn{2}{c}{(mJy)} & & (Kpc)& W/Hz &  W/Hz & W/Hz & \\
\hline	 
{\bf \object{J1504+689}} & 15 04 12.781 & 68 56 12.75 &  451 &  3.4 & 88&4 & 72&1 & 0.318$^a$ &  1160 & 26.35 & 25.64 & 25.55 & II-QSS\\
\object{J1523+636} & 15 23 45.900 & 63 39 23.78 &  676 &  3.5 & 18&3e & 14&5  & 0.204$^a$ &  899 & 26.12 & 24.55 & 24.45 & II\\
\object{J1530+824} & 15 30 56.110 & 82 27 21.02 &  180 &  6.6 & 40&6  & 27&2e & 0.021$^a$ &  236 & 23.55 & 22.90 & 22.73 & I \\
\object{J1536+843} & 15 36 57.335 & 84 23 10.42 &  375 &  8.0 &  4&9  &  3&3  & 0.201$^b$ & 2033 & 25.85 & 23.97 & 23.80 & II\\
\object{J1557+706} & 15 57 30.190 & 70 41 20.79 & 1800 & 11.3 & 25&2  & 31&3  & 0.026$^a$ &  490 & 24.72 & 22.87 & 22.96 & I \\
\object{J1632+825} & 16 32 31.630 & 82 32 16.28 & 2200 & 66.0 &428&2  &286&0  & 0.023$^a$ & 2544 & 24.70 & 23.99 & 23.82 & I \\
\object{J1650+815} & 16 50 58.686 & 81 34 28.11 &  313 &  5.6 & 44&4  & 32&8  & 0.038$^a$ &  348 & 24.30 & 23.45 & 23.32 & I \\
\object{J1732+714} & 17 32 33.001 & 71 24 10.50 &  616 &  4.0 & 35&1e &  9&5e & 0.059$^a$ &  372 & 24.98 & 23.73 & 23.16 & I \\
\object{J1733+707} & 17 33 12.525 & 70 46 30.36 &  240 &  4.5 & 10&6e &  8&1  & 0.041$^a$ &  299 & 24.24 & 22.89 & 22.77 & I \\
\object{J1743+712} & 17 43 17.681 & 71 12 53.98 &  142 &  4.0 &  9&2  & 13&1  & ---       &    --- &   --- &   --- &   --- & II\\
{\bf \object{J1745+712}} & 17 45 43.573 & 71 15 48.55 &  883 &  4.4 & 25&9e & 10&9  & 0.216$^a$ & 1176 & 26.29 & 24.76 & 24.38 & II\\
\object{J1751+680} & 17 51 19.629 & 68 04 43.05 &  153 &  8.1 & 11&8e &  9&0  & 0.079     &   978 & 24.63 & 23.52 & 23.40 & I\\
\object{J1754+626} & 17 54 50.310 & 62 38 41.96 &  991 & 16.2 &  5&0  &  5&6  & 0.028$^a$ &  754 & 24.53 & 22.23 & 22.28 & I\\
\object{J1800+717} & 18 00 42.622 & 71 44 41.99 &  144 &  4.2 &  1&4  &  0&9  &   ---     &    --- &   --- &   --- &   --- & II\\
{\bf \object{J1835+665}} & 18 35 07.338 & 66 35 00.02 &  136 &  4.4 &   2&9 &  1&6  & 0.354?   & 1594 & 25.93 & 24.26 & 24.00 & II\\
\object{J1835+620} & 18 35 10.405 & 62 04 07.42 &  800 &  3.9 &   2&0e&  1&7  & 0.518     & 1688 & 27.05 & 24.45 & 24.38 & II\\
{\bf \object{J1844+653}} & 18 44 07.443 & 65 22 03.07 &  104 &  7.5 &   0&8 &  1&9  & 0.197    & 1881 & 25.28 & 23.16 & 23.54 & II\\
\object{J1845+818} & 18 45 15.836 & 81 49 30.98 &  596 &  4.4 &   5&2 &  6&9  & 0.119     &  750 & 25.58 & 23.52 & 23.65 & II\\
\object{J1847+707} & 18 47 34.912 & 70 44 00.64 &  226 &  3.8 &  31&2e&  8&1e & 0.043     &  265 & 24.26 & 23.40 & 22.82 & I \\
\object{J1850+645} & 18 50 45.871 & 64 30 24.68 &  154 &  5.6 &  10&6e&  4&0e & 0.080     &  683 & 24.64 & 23.48 & 23.06 & I\\
{\bf \object{J1853+800}} & 18 53 52.077 & 80 02 50.46 &  155 &  5.6 &   3&4e&  2&1  & 0.214$^a$ & 1486 & 25.53 & 23.87 & 23.66 & II\\
{\bf \object{J1918+742}} & 19 18 34.885 & 74 15 05.05 &  570 &  6.6 &  26&1 &  8&8  & 0.194    & 1636 & 26.00 & 24.66 & 24.19 & II\\
{\bf \object{J1951+706}} & 19 51 40.825 & 70 37 39.99 &  100 &  5.2 &   3&4 &  6&0  & 0.550     & 2303 & 26.20 & 24.74 & 24.98 & II\\
\object{J2016+608} & 20 16 18.630 & 60 53 57.49 &  332 &  3.0 &   2&3e&  2&2  & 0.121    &  519 & 25.35 & 23.19 & 23.17 & II\\
{\bf \object{J2035+680}} & 20 35 16.549 & 68 05 41.60 &  156 & 11.5 &  10&6 & 15&5  & 0.133   & 2143 & 25.10 & 23.93 & 24.10 & I\\
\object{J2042+751}$^c$ & 20 42 37.180 & 75 08 02.52 & 1805 & 10.6 & 184&0 & 328&0 & 0.104$^a$ & 1617 & 25.94 & 24.95 & 25.20 & II-QSS\\
{\bf \object{J2059+627}} & 20 59 09.560 & 62 47 44.11 &  113 &  4.7 &  3& 0 &  4&0  & 0.267   & 1444 & 25.59 & 24.01 & 24.14 & II? \\
{\bf \object{J2103+649}} & 21 03 13.868 & 64 56 55.26 &  123 &  4.8 &  9& 8 &  7&3e & 0.215   & 1279 & 25.43 & 24.33 & 24.20 & II \\
\object{J2111+630} & 21 11 20.268 & 63 00 06.17 &285&7.0&\multicolumn{2}{c}{---}& 1&0 &  --- & --- &   --- &   --- &   --- & II \\
\object{J2114+820} & 21 14 01.179 & 82 04 48.28 &  483 &  6.0 &141&1 &140&9   & 0.085     &  772 & 25.19 & 24.66 & 24.66 & I \\
\object{J2128+603} & 21 28 02.634 & 60 21 07.96 &  258 &  5.9 &  4&5e&  6&4   & 0.072?    &  656 & 24.77 & 23.02 & 23.17 & II \\
\object{J2138+831} & 21 38 42.266 & 83 06 49.21 &  305 &  5.1 & 30&2 & 13&8   & 0.135     &  962 & 25.41 & 24.40 & 24.06 & ? \\
\object{J2145+819} & 21 45 29.887 & 81 54 54.22 &  400 & 18.7 & 12&0 &  8&2   & 0.146$^b$ & 3739 & 25.59 & 24.07 & 23.90 & II \\
\object{J2157+664} & 21 57 02.572&66 26 10.24& 1070 &5.0&\multicolumn{2}{c}{---}&34&6& 0.057?& 451 &  25.19 & --- &  23.70 & ? \\
\object{J2204+783} & 22 04 09.225 & 78 22 46.92 &  251 &  4.0 &  3&4 &  5&4   & 0.115     &  663 & 25.18 & 23.31 & 23.51 & II \\
\object{J2209+727} & 22 09 33.780 &72 45 58.36 &248 & 3.7&\multicolumn{2}{c}{---}& 0&4 & 0.201 &940 & 25.67 &--- & 22.88 & II \\
{\bf \object{J2242+622}} & 22 42 32.133 & 62 12 17.53 &  287 &  4.2 &  1&6 &  1&9   & 0.188?& 1018 & 25.68 & 23.42 & 23.50 & II \\
\object{J2247+633} & 22 47 29.714 & 63 21 15.55 &  477 &  4.4 &  6&7e&  4&2   &   ---     &    --- &   --- &   --- &   --- & I \\
\object{J2250+729} & 22 50 43.621 &72 56 16.19 &653 &3.8&\multicolumn{2}{c}{---}& 0&9e & --- &  --- &  --- &   --- &   --- & II \\
\object{J2255+645} & 22 55 29.943 & 64 30 06.86 &  392 &  4.0 &  4&9 &  7&3   &   ---     &    --- &   --- &   --- &   --- & II \\
\object{J2307+640} & 23 07 58.533 & 64 01 39.22 &  171 &  4.6 &  7&3 &  4&3   &   ---     &    --- &   --- &   --- &   --- & II \\
\object{J2340+621} & 23 40 56.435 & 62 10 45.09 &  185 &  4.1 &  2&0 &  2&6   &   ---     &    --- &   --- &   --- &   --- & I \\
\hline 
\end{tabular}
\parbox{16cm}{
$^{\ast}$ I and II stand for FR I and FR II type radio galaxies, respectively\\
$^a$ Redshift from the NASA Extragalactic Database\\
$^b$ Redshift from Schoenmakers (\cite{arnotesis})\\
$^c$ Core flux densities taken from Pearson et al. (\cite{pearson2}).\\
Label ``e'' by the core flux density means extended core (see text). \\
New giant radio galaxies are written in boldface.
}
\end{scriptsize}
\end{table*}

\begin{table*}[t]
\caption[]{Map parameters}
\label{parameters}
\begin{scriptsize}
\begin{tabular}{llr@{$\times$}lrlr@{$\times$}lrl}
Name    & Observations$^{a}$  & \multicolumn{4}{c}{1.4 GHz} & \multicolumn{4}{c}{4.9 GHz} \\
    &   & \multicolumn{3}{c}{Beam} & 1st contour & \multicolumn{3}{c}{Beam} & 1st contour  \\      
    &   & $\arcsec$ & $\arcsec$ &P.A.& mJy/beam & $\arcsec$ & $\arcsec$ & P.A.& mJy/beam  \\
\hline	 
J0109+731  & j,k      &  5.9 &  3.7 & -39.6 & 0.86 & 1.7 & 1.7 &           & 0.39 \\       
J0153+712  & j,k,l    & 10.7 &  9.6 & -31.9 & 1.0  & \multicolumn{3}{c}{--}& N.S. \\ 
J0317+769  & b,c,d,g,l& 22.0 & 12.1 & -68.0 & 0.7  & 3.7 & 2.9 & -88.3     & 0.15 \\ 
J0318+684  & a,b,c &\multicolumn{3}{c}{--}&N.S.$^b$ &\multicolumn{3}{c}{--}& N.S. \\
J0342+636  & b,f,k    & 10.6 &  9.7 &  68.8 & 0.45 & 1.5 & 1.5 &           & 0.16 \\
J0430+773  & b,c,f,g  & 10.6 & 10.2 &  82.9 & 0.6  & 3.8 & 3.0 & -68.9     & 0.18 \\ 
J0455+603  & b,c,f    &  5.9 &  4.7 &  46.1 & 0.14 & \multicolumn{3}{c}{--}& N.S. \\
J0502+670  & b,c,f,k  & 10.7 &  9.9 &  89.2 & 0.2  & 3.3 & 3.1 & -64.6     & 0.1  \\
J0508+609  & a,b,c,g  & 15.0 & 10.2 & -74.5 & 0.2  & 3.3 & 2.8 &  63.0     & 0.2  \\
J0519+702  & a,b,c,g  & 13.4 & 10.4 &  82.4 & 0.4  & 4.4 & 2.8 & -56.1     & 0.1  \\ 
J0525+718  & a,b,c,g  & 13.3 & 10.0 &  83.1 & 0.4  & \multicolumn{3}{c}{--}& N.S. \\ 
J0531+677  & b,c,d,g  & 25.0 & 25.0 &       & 0.8  & 1.6 & 1.6 &           & 0.17 \\
J0546+633  & a,b,c,g  & 15.4 & 10.5 & -67.8 & 0.25 & 4.2 & 3.1 &  72.6     & 0.1  \\ 
J0559+607  & b,c,f    & 11.2 &  9.3 &  59.5 & 0.2  & \multicolumn{3}{c}{--}& N.S. \\ 
J0607+612  & a,b,c,k  & 15.2 & 11.5 & -68.8 & 0.4  & 4.1 & 3.0 &  59.0     & 0.1  \\ 
J0624+630  & b,c,f,g  & 11.2 &  9.5 &  66.0 & 0.3  & 3.9 & 3.3 &  69.3     & 0.1  \\ 
J0633+721  & a,b,c,g  & 13.2 & 10.9 & -82.3 & 0.3  & 4.6 & 2.7 &  44.0     & 0.1  \\ 
J0654+733  & a,b,c    & 16.2 & 12.5 &  60.6 & 0.4  & \multicolumn{3}{c}{--}& N.S. \\
J0750+656  & d,h,k    &  9.2 &  4.8 & -80.9 & 0.5  & 1.7 & 1.7 &           & 0.15 \\ 
J0757+826  & d,h,m    & 22.0 & 22.0 &       & 5.0  & 6.1 & 6.1 &           & 0.3  \\
J0803+669  & b,c,f    &  6.7 &  4.7 &  58.9 & 0.35 & \multicolumn{3}{c}{--}& N.S. \\ 
J0807+740  & a,b,c    & 13.0 & 11.3 & -30.2 & 0.3  & \multicolumn{3}{c}{--}& N.S. \\ 
J0819+756  & a,b,c,g  & 13.9 & 12.6 &  13.2 & 0.3  & 7.7 & 7.7 &           & 0.4  \\ 
J0825+693  & a,b,c,g &\multicolumn{3}{c}{--}& N.S. & \multicolumn{3}{c}{--}& N.S. \\
J0828+632  & d,h,k    &  9.8 &  6.3 & -72.1 & 0.4  & 1.7 & 1.7 &           & 0.15 \\
J0856+663  & b,c,f    & 13.4 & 13.4 &       & 0.6  & \multicolumn{3}{c}{--}& N.S. \\ 
J0926+653  & d,h,m    & 26.0 & 26.0 &       & 1.3  & 6.1 & 6.1 &           & 0.15 \\ 
J0926+610  & d,k      & 12.3 &  4.8 & -63.8 & 0.7  & 1.7 & 1.7 &           & 0.14 \\ 
J0939+740  & b,c,f    & 12.0 & 12.0 &       & 0.6  & 7.0 & 7.0 &           & 0.15 \\ 
J0949+732  & --       &\multicolumn{3}{c}{--}& N.S. & \multicolumn{3}{c}{--}& N.S. \\
J1015+683  & d,g,h,i  & 12.2 &  4.8 & -58.0 & 0.6  & 4.8 & 3.6 & -11.7     & 0.13 \\
J1036+677  & d,e,h,i  & 12.4 &  4.8 & -58.1 & 0.45 & 5.5 & 3.2 & -59.7     & 0.1  \\ 
J1124+749  & f,h,m    & 25.0 & 25.0 &       & 0.5  & 6.8 & 6.8 &           & 0.15 \\
J1137+613  & d,m      &  9.0 &  9.0 &       & 1.0  & 7.3 & 7.3 &           & 0.2  \\
J1211+743  & f,g,h,i  & 12.0 & 12.0 &       & 0.4  & 4.4 & 3.8 & -32.9     & 0.1  \\ 
J1216+674  & d,m      &  8.5 &  8.5 &       & 0.5  & \multicolumn{3}{c}{--}& N.S. \\
J1220+636  & d,l,m    &  8.9 &  8.9 &       & 0.9  & 7.3 & 7.3 &           & 0.16 \\
J1247+673  & d,e,l    & 26.0 & 26.0 &       & 1.0  & \multicolumn{3}{c}{--}& N.S. \\
J1251+756  & f,h,m    & 12.6 & 11.3 &  39.0 & 0.3  & 6.8 & 6.8 &           & 0.1  \\ 
J1251+787  & d,e,h,i  & 24.0 & 24.0 &       & 0.5  & 5.4 & 3.4 & -29.6     & 0.1  \\ 
J1313+696  & d,e      &  8.0 &  8.0 &       & 0.9  & 2.5 & 2.5 &           & 0.5  \\ 
J1410+633  & d,m      & 10.0 & 10.0 &       & 0.6  & 8.4 & 8.4 &           & 0.3  \\ 
\hline 
\multicolumn{10}{l}
{\parbox{11cm}{$^{a}$ The codes in this column refer to those in Table~\ref{obs}\\
$^b$ N.S. stands for not shown sources in Fig.~\ref{mapas}}}
\end{tabular}
%\parbox{11cm}{
%$^a$ The codes in this column refer to those in Table~\ref{obs} \\
%$^b$ N.S. stands for not shown sources in Fig.~\ref{mapas}.
%}
\end{scriptsize}
\end{table*}

\begin{table*}[t]
\addtocounter{table}{-1}
\caption[]{({\em cont.}) Map parameters}
%\label{parameters}
\begin{scriptsize}
\begin{tabular}{llr@{$\times$}lrlr@{$\times$}lrl}
Name    & Observations$^a$ & \multicolumn{4}{c}{1.4 GHz} & \multicolumn{4}{c}{4.9 GHz} \\
   &    & \multicolumn{3}{c}{Beam} & 1st contour & \multicolumn{3}{c}{Beam} & 1st contour  \\      
   &    & $\arcsec$ & $\arcsec$ &P.A.& mJy/beam & $\arcsec$ & $\arcsec$ & P.A.& mJy/beam  \\
\hline	 
J1504+689  & d,k      & 10.0 & 10.0 &       & 0.8  & 1.8 & 1.8 &           & 0.3  \\
J1523+636  & d,e      & 11.0 & 11.0 &       & 0.8  & 3.1 & 3.1 &           & 0.22 \\
J1530+824  & f,h,m    & 25.0 & 25.0 &       & 1.0  & 6.7 & 6.7 &           & 0.15 \\
J1536+843  & f,h,m    &  9.8 &  5.2 &  80.6 & 0.2  & 6.6 & 6.6 &           & 0.15 \\ 
J1557+706  & d,g,l    & 10.9 &  8.9 &  77.0 & 1.2  & 2.0 & 2.0 &           & 0.25 \\
J1632+825  & f,g,h,i  & 13.9 &  9.5 &  69.9 & 1.0  & 3.3 & 3.3 &           & 0.2  \\
J1650+815  & f,h,m    & 25.0 & 25.0 &       & 0.63 & 6.6 & 6.6 &           & 0.2  \\
J1732+714  & d,e      &  9.0 &  4.8 & -60.6 & 0.3  & 2.8 & 1.4 & -78.0     & 0.2 \\
J1733+707  & d,m      &  7.9 &  6.7 & -3.1  & 0.57 & 6.6 & 6.6 &           & 0.14 \\
J1743+712  & d,g,l    &  7.4 &  5.0 & -66.0 & 0.4  & 1.8 & 1.8 &           & 0.16 \\
J1745+712  & d,g,l    &  8.0 &  5.7 & -78.5 & 0.55 & 1.8 & 1.8 &           & 0.2  \\
J1751+680  & d,k,l    &  6.9 &  4.7 & -64.2 & 0.35 & 1.8 & 1.4 & -23.2     & 0.12 \\
J1754+626  & d,g,l    & 24.0 & 24.0 &       & 0.8  & 1.8 & 1.8 &           & 0.15 \\
J1800+717  & d,k      &  7.1 &  4.7 & -61.8 & 0.45 & \multicolumn{3}{c}{--}& N.S.$^b$ \\ 
J1835+665  & d,m      &  6.8 &  4.7 & -58.4 & 0.35 & 6.8 & 6.8 &           & 0.2  \\
J1835+620  & d,e,l,m  & \multicolumn{3}{c}{--}&N.S.& \multicolumn{3}{c}{--}& N.S. \\
J1844+653  & d,l,m    & 24.0 & 24.0 &       & 1    & 7.0 & 7.0 &           & 0.14 \\
J1845+818  & f,h,m    & 13.8 &  9.4 &  87.7 & 0.5  & 6.3 & 6.3 &           & 0.14 \\ 
J1847+707  & d,e,l    & 12.2 & 10.1 & -87.6 & 0.5  & 2.2 & 1.4 & -60.1     & 0.15 \\
J1850+645  & d,g,l    & 25.0 & 25.0 &       & 1.0  & \multicolumn{3}{c}{--}& N.S. \\
J1853+800  & f,g,h,i  & 14.2 &  9.6 &  86.7 & 0.4  & 5.0 & 3.4 &  74.6     & 0.15 \\
J1918+742  & j,k,l    & 13.0 & 13.0 &       & 0.8  & \multicolumn{3}{c}{--}& N.S. \\
J1951+706  & j,l,m    & 12.5 & 12.5 &       & 0.35 & 6.8 & 6.8 &           & 0.2  \\ 
J2016+608  & j,k      &  6.2 &  4.9 & -50.6 & 0.4  & 1.7 & 1.7 &           & 0.15 \\
J2035+680  & j,k,l    & 22.0 & 22.0 &       & 0.65 & 1.7 & 1.7 &           & 0.15 \\
J2042+751  & --       &\multicolumn{3}{c}{--}& N.S. & \multicolumn{3}{c}{--}& N.S. \\
J2059+627  & j,m      &  5.5 &  5.5 &       & 0.5  & 7.0 & 7.0 &           & 0.15 \\
J2103+649  & j,l,m    & 11.7 & 10.2 &  84.3 & 0.3  & 6.7 & 6.7 &           & 0.16 \\ 
J2111+630  & j,k,l    & 22.0 & 22.0 &       & 1.0  & \multicolumn{3}{c}{--}& N.S. \\
J2114+820 &a,b,c,g,j,k& 12.8 &  9.7 &  11.1 & 0.4  & 2.9 & 2.3 & -65.5     & 0.12 \\
J2128+603  & j,k      &  5.4 &  3.8 & -51.5 & 0.25 & 1.7 & 1.7 &           & 0.15 \\ 
J2138+831  & a,b,c,g  & 12.8 &  9.6 &  16.7 & 0.3  & 6.5 & 3.4 &  40.8     & 0.2  \\ 
J2145+819  & a,b,c    & 12.7 &  9.8 &  19.0 & 0.4  & \multicolumn{3}{c}{--}& N.S. \\ 
J2157+664  & j,k,l,m  & 11.6 & 10.1 &  87.5 & 0.53 & 4.1 & 3.2 &  88.5     & 0.2  \\
J2204+783  & a,b,c,g  & 13.0 & 10.2 &  34.1 & 0.3  & 4.1 & 3.0 &  45.1     & 0.14 \\
J2209+727  & j,k      &  7.4 &  5.2 & -57.7 & 0.3  & 1.8 & 1.8 &           & 0.22 \\ 
J2242+622  & j,k      &  5.1 &  3.8 & -36.9 & 0.25 & 1.6 & 1.6 &           & 0.14 \\ 
J2247+633  & j,k      &  6.3 &  5.0 & -45.1 & 0.4  & 1.6 & 1.6 &           & 0.1  \\ 
J2250+729  & j,k      &  7.2 &  4.9 & -51.7 & 0.4  & 1.8 & 1.8 &           & 0.2  \\ 
J2255+645  & j,k      &  6.8 &  4.8 & -60.2 & 1.0  & \multicolumn{3}{c}{--}& N.S. \\ 
J2307+640  & j,k      &  5.7 &  3.7 & -53.9 & 0.3  & \multicolumn{3}{c}{--}& N.S. \\ 
J2340+621  & j,k      &  5.5 &  3.8 & -50.1 & 0.25 & 1.6 & 1.6 &           & 0.15 \\
\hline 
\multicolumn{10}{l}
{\parbox{11cm}{$^{a}$ The codes in this column refer to those in Table~\ref{obs}\\
$^b$ N.S. stands for not shown sources in Fig.~\ref{mapas}}}
\end{tabular}
%\parbox{11cm}{
%$^a$ The codes in this column refer to those in Table~\ref{obs} \\
%$^b$ N.S. stands for not shown sources in Fig.~\ref{mapas}.
%}
\end{scriptsize}
\end{table*}

\section{Notes on individual sources}

Selected radio maps of 79 objects in our sample are displayed in
Fig.~\ref{mapas}. We do not present maps of the radio sources
J0318+684, J0825+693 and J1835+620, which have been discussed in
separate papers (Schoenmakers et al. \cite{arno1}, Lara et
al. \cite{lara3}, Lara et al. \cite{lara1}, respectively). We neither
present maps of the sources which we have not observed: the giant
radio galaxy J0949+732 (4C+73.08; J\"agers \cite{jagers}; Mack
\cite{mack}; Leahy et al. \cite{leahy}) and the giant radio 
quasar J2042+751 (4C74.26; Riley et al. \cite{riley}). For a few sources, 
images at 4.9 GHz which do
not add relevant information about the source structure are not shown
for brevity.  The observations of each source of the sample and the
parameters of each map in Fig.~\ref{mapas} (size and position angle
(P.A.) of the convolution beam and the brightness level of the first
contour) are displayed in Table~\ref{parameters}.  Unless differently
commented, all the radio structure observed in the NVSS maps is also
found in the presented images at 1.4 GHz.  To ease the comparison of
the different radio sources, we have tried to always use the same map
scale at 1.4 GHz, except for those sources whose size is too large to
properly fit in a single page, for which the scale has been doubled or
tripled (marked with a ``1/2'' or ``1/3'' on the corresponding
map). At 4.9 GHz the map scale has been optimized to show the details
of the radio structure.

In the following, we briefly describe those radio sources in the
sample which have peculiar properties and/or morphologies somehow
different from the typical FR I or II double structure.

{\bf J0109+731 (3C\,31.1)}: an FR II type Broad Line Radio Galaxy (BLRG) 
(eg. van Breugel \&
J\"agers \cite{vanbreugel}). The NE lobe is much shorter than the SW
one and in addition it has a prominent hotspot at its extreme, which
suggests that the interaction with the external medium plays an
important role in the source asymmetry. A jet in the SW direction is
observed at 1.4 and at 4.9 GHz.

{\bf J0153+712}: a low power but core dominated radio galaxy, 
classified as a BL-Lac candidate by Marcha et al. (\cite{marcha}). 
It is strongly asymmetric, with a wide jet directed in W-NW direction, 
with kinks and bends,
which fades to the noise level at about 5$\arcmin$ from the core in
our 1.4 GHz map. We also detect a component in the opposite direction,
most probably from a counter-jet. At 4.9 GHz we only detect the strong
core component.

{\bf J0317+769}: presents a peculiar morphology consisting of a core
and two FR I type symmetric and prominent jets, which end in two FR II
type lobes (better seen in the 4.9 GHz map). From the SW lobe, a faint
and extended tail rises directed in S direction (1.4 GHz map). This
tail is completely resolved out in our 4.9 GHz map.

{\bf J0455+603}: a peculiar radio source with a well defined core
embedded in a low brightness halo-like structure in NW-SE direction.
Our 1.4 GHz map shows marginal evidence of a jet in NW direction. 
No redshift information is available for this radio source.

{\bf J0508+609}: a low power FR I type giant radio galaxy with a
prominent flat spectrum core and two opposite jets directed in SE-NW
direction.  Both jets present large opening angles.  The northern jet
can be followed at larger distances from the core than the southern
one, which is resolved out beyond 1.25$\arcmin$ from the core in our
map at 1.4 GHz, but clearly seen in the NVSS map up to 6$\arcmin$ from
the core.  At 4.9 GHz we only detect the core component, a 13$\arcsec$
jet in SE direction and marginal evidence of the counter-jet on the
opposite side.

{\bf J0525+718}: a low power radio galaxy with two symmetric jets in
the E-W direction. The western jet shows a 90$^{\circ}$ bend in
projection towards the north. It is not clear if this bend is somehow
related to a compact source seen in the same field.  The total flux
density of J0525+718 at 1.4 GHz is below the 100 mJy limit after the
subtraction of the flux density from this compact source, but we have
kept it in the final sample.  At 4.9 GHz we only detect a weak core
component and a very faint jet and counter-jet emission (map not
shown).

{\bf J0531+677}: our 4.9 GHz map shows a weak core and two symmetric
jets in the NE-SW direction, typical of an FR I structure. What is
peculiar in this radio source is the existence of a very faint and
extended tail directed towards the south, similar to that previously
described in J0317+769. A similar tail might be present in the other
jet as in wide angle tail radio galaxies, but the emission quickly fades 
below the noise level. It is the nearest object in the sample, at z=0.017.

{\bf J0546+633}: an asymmetric FR I type radio galaxy with a NE-SW
orientation. The jet and counter-jet are prominent in our maps and
well collimated. While the SW jet fades away as it separates from the
core, the NE jet is shorter and presents a strong bending
backwards. This is another example of a radio galaxy with an
asymmetric structure, possibly resulting from 
different degrees
of interaction of the oppositely directed jets with the external medium
and/or the projection effects.

{\bf J0607+612}: a radio galaxy with lobes in the NE-SW direction. Its
projected linear size is 1.47 Mpc (GRG). At 4.9 GHz we can clearly
follow the jet directed towards the NE lobe. The structure of
J0607+612 cannot be easily classified as an FR I or FR II type.

{\bf J0624+630}: a low power FR I radio source oriented in N-S
direction with an S-shaped morphology. The N-jet presents a strong bend
at 0.7$\arcmin$ from the core. The core component is clearly
identified in the 4.9 GHz map.

{\bf J0654+733}: a 2 Mpc long FR II radio galaxy directed in the NE-SW
direction, with prominent hotspots at the lobe extremes. The core is
displaced towards the SW with respect to the center of
symmetry of the source. A background compact source, most probably
unrelated, is located close to the northern lobe.

{\bf 0750+656}: the most distant object of our sample, at z=0.747. It
is a GRG with a linear projected size of 1.8 Mpc. It presents a
prominent core and a FR II type radio structure in NW-SE direction.

{\bf J0807+740}: a giant low power radio galaxy. At 1.4 GHz, this
peculiar radio source presents a compact core component and a weak and
extended halo-like emission elongated in the E-W direction. There is
no evidence of jets or hotspots in our maps. At 4.9 GHz we only detect
the core component. This object could be a relic FR II radio galaxy, where
hotspot regions are no more present and extended lobes are still detected at
1.4 GHz.

{\bf J0819+756}: an FR II type GRG (linear size of 2.2 Mpc) oriented
in NE-SW direction. It presents a strong inverted spectrum core
($\alpha_{1.4}^{4.9}=0.5$; Flux density $\propto\nu^{\alpha}$). Both lobes
harbour a double hotspot structure at their extremes.  The source
beyond the NE lobe is an unrelated background object, as suggested by
its structure and its optical identification.

{\bf J1015+683}: presents a very complex and distorted structure
oriented in the E-W direction resembling, mostly at 4.9 GHz, two double
radio galaxies closely seen in projection, one slightly above the
other. This is supported by the optical image which shows two nearby
galaxies in the field identifed with radio components (Paper II). 
Redshift and core data on Table~\ref{sample} refer to the northern feature.

{\bf J1137+613}: a symmetric radio galaxy of FR II type morphology with 
prominent hotspots at the end of the lobes. There is evidence of a strong
backflow from both radio lobes which is deflected in opposite
directions perpedicularly to the axis defined by the jets.

{\bf J1211+743}: a GRG (linear size of 1.2 Mpc) which presents a
peculiar structure, not clearly discriminated between FR I or II types. 
There is a prominent jet in the NW direction, 
showing blobs of
emission and a bent structure. A faint counter-jet is also
detected. No strong hotspots are observed in the lobes.

{\bf J1247+673}: a known Gigahertz Peaked Spectrum radio galaxy with a
FR II type morphology (de Vries et al. \cite{devries}), also a member of
the class of giant radio galaxies (linear size of 1.94 Mpc). It clearly 
contrasts with respect to the rest of the sample in its core dominance 
($67\%$ of the source emission at 1.4 GHz comes from the core, which is the 
only feature detected at 4.9 GHz).  

{\bf J1251+787}: one of the most extended radio galaxies in our sample
(19.5$\arcmin$; redshift is not available for this source). It
presents a FR I type morphology, with two irregular S-shaped and well
collimated jets. They present several blobs of emission and bends
until they fall below the noise level of our images.

{\bf J1313+696 (4C+69.15)}: an FR II type radio galaxy oriented in
SE-NW direction. With a projected linear size of 1.06 Mpc, it belongs
to the class of giants.  At 1.4 GHz the emission from the lobes form a
continuous bridge along the entire source. At 4.9 GHz only the core
and the lobe extremes are detected. This case emphasizes the
importance of the observations made at 4.9 GHz to discriminate the
core component, needed to identify the associated galaxy.

{\bf J1504+689 (4C+69.18)}: one of the few quasars in the sample,
presents a typical FR II type morphology with a bright flat-spectrum
core and two lobes with prominent hotspots. With a projected linear size
of 1.16 Mpc, it is a giant radio quasar. There is an unrelated strong
compact source at $2.16\arcmin$ from the core, in SW direction, which
was misidentified with a component of J1504+689 by Reid et
al. (\cite{reid}). 

{\bf J1557+706 (4C+70.19)}: it presents an FR I type morphology in the
N-S direction. The northern jet bends by 180$^{\circ}$ towards the
south, becoming diffuse and extended beyond the bend. The southern jet
ends in an extended lobe-like region. At 4.9 GHz we only detect a
compact core and the beginning of two rather symmetric jets, separated
from the core by symmetric gaps.

{\bf J1632+825 (NGC~6251)}: a very well studied radio galaxy
(eg. Perley et al. \cite{perley1}), and the most extended one in our
sample (66$\arcmin \equiv 2.54$ Mpc). Our observations show only a
prominent and well collimated one-sided jet in the NW direction, with no
evidence of the counter-jet emission. Total angular size and total
flux density in Table~\ref{sample} are from Perley et
al. (\cite{perley1}).

{\bf J1650+815}: a peculiar low brightness asymmetric radio galaxy.
The core is clearly identified at 4.9 GHz, and appears displaced
towards the north with respect to the center of the radio
structure. There is evidence of a jet departing from the core in S-SW
direction, and two extended and diffuse lobes.

{\bf J1732+714}: an FR I type radio galaxy with two symmetric jets
departing from a central core. The jets end in two extended lobes
without strong hotspots. It presents a bridge of low brightness
emission southward of the jets and in the same direction, very much like
the relic emission observed in 3C\,338 (Giovannini et al. \cite{giovannini}).

{\bf J1745+712 (4C+71.17)}: at 1.4 GHz it shows an FR II type
morphology in the E-W direction with a dominant bright emission from the core
region. Higher resolution observations at 4.9 GHz show a small and prominent 
structure ($\sim 15\arcsec$) with a core and a two symmetric jet components,
which might be the result of an episode of enhanced activity, as in the 
case of J1835+620 (Lara et al. \cite{lara1}).

{\bf J1751+680}: a wide angle tail radio source with detached lobes
displaced towards the east. The southern jet presents a sharp bend of
$\sim 90^{\circ}$. An unrelated double radio source appears confused
with the southern lobe when observed at low angular resolution
(e.g. in the NVSS map).

{\bf J1754+626 (NGC~6512)}: another wide angle tail source, with an
extension of low brightness emission in the southern direction. The
radio structure is very complex, with sharp bends and kinks, most
likely produced by the interaction with the external medium. At 4.9 GHz 
we observe the core component and two jets in the E-W direction. The eastern
jet beds sharply towards the west, most probably due to the existence
of a strong intergalactic wind.

{\bf J1918+742}: a GRG with projected linear size of 1.64 Mpc. It
presents an FR II type morphology in the E-W direction, with the core
strongly displaced towards the east from the center of symmetry of the
source. At 4.9 GHz we only detect the core and the hotspot in the
western lobe.

{\bf J2035+680}: an FR I type radio galaxy in the N-S direction with two
extended and diffuse lobes. The northern jet presents a prominent blob
of emission at $2\arcmin$ from the core.  The angular size is
$11.5\arcmin$, which correspond to 2.14 Mpc in projected linear size
for a redshift z=0.133 (uncertain). There is a strong and compact
radio source almost superposed to the southern lobe, which we consider
unrelated to J2035+680.

{\bf J2111+630}: a peculiar source with an FR II type morphology in the
NW-SE direction. At 1.4 GHz, the lobes appear extended with signs of
backflow, but there are not prominent hotspots. At 4.9 GHz we only
detect a weak core (flux density $\sim$ 1 mJy). We could not determine
the redshift of the associated galaxy.

{\bf J2114+820}: presents an FR I type structure, with S-shaped
extended lobes and a strong and variable flat spectrum core (variability 
observed in our data at different epochs). There is
a prominent jet in the NW direction with several blobs of emission and
with a very strong widening at about $\sim 20\arcsec$ from the core. A
weaker counter-jet is observed on the opposite side.  Its optical
spectrum presents prominent broad emission lines (Stickel et
al. \cite{stickel}; Paper II), which according with current
unification schemes of radio loud Active Galactic Nuclei (e.g. Urry \&
Padovani \cite{urry}), is in contradiction to its classification as
an FR I radio galaxy (see also Lara et al. \cite{lara2}).

{\bf J2138+831}: a strongly asymmetric radio galaxy in the cluster \object{Abell
2387}. While the eastern lobe bends southwards forming a long tail of
FR I type, the western lobe resembles those found in FR II type radio
sources. At 4.9 GHz we observe how the jet directed towards the west
loses its collimation at $\sim 30\arcsec$ from the core.

{\bf J2145+819}: an FR II type radio galaxy with a N-S orientation. It
is the intrinsically largest radio galaxy in our sample (a giant among
the giants, with a projected linear size of 3.74 Mpc).  The northern
lobe presents an almost perfect conical shape with a prominent hotspot
at its extreme (see Lara et al. \cite{lara3}). Palma et
al. (\cite{palma}) present a detailed study of this radio source.

{\bf J2157+664 (4C66.24)}: a very peculiar and asymmetric radio
source. At first sight (1.4 GHz observations) it seems to have two
typical radio lobes in the E-W direction, one of them with a strong
hotspot. However, observations at 4.9 GHz and spectral index
considerations show that the core is hosted by the western ``lobe'',
and that a jet directed towards the west interacts at a very short
distance from the core with the external medium, producing a very
strong bow-shock. On the opposite side, there is a long jet which
seems to bend in an almost closed loop, forming a long tail of low
brightness emission towards the south (observed at 1.4 GHz).

{\bf J2209+727}: an FR II type radio galaxy in the NW-SE direction. The core
is very weak and appears confused with the lobe emission at 1.4 GHz. At
4.9 GHz we can clearly identify the weak core (flux density $400 \mu$Jy)
and the hotspots at the extremes of the lobes. There are two nearby
radio sources in the same field, most probably unrelated with J2209+727.

{\bf J2340+621}: a peculiar radio source extended in the E-W direction,
with two collimated S-shaped jets. In both jets there are prominent
blobs of emission at $\sim 22\arcsec$ (E-jet) and $\sim 30\arcsec$
(W-jet) from the core, which suggest different phases in the core
activity, as in the case of J1745+712 or J1835+620 (Lara et
al. \cite{lara1}).  This radio source is located at a very low
galactic latitude, 0.4$^{\circ}$. We detected its optical counterpart,
although its extragalactic nature could not be confirmed (Paper II).

\section{Discussion}

We discuss in this section possible biases introduced in our sample
due to the criteria adopted for the sample selection and to the sensitivity
limitations of the NVSS. A detailed discussion about the implications
on the radio source population derived from the sample is left for a
forthcoming paper (Paper III).

\begin{figure}[b]
\vspace{8cm}
\includegraphics{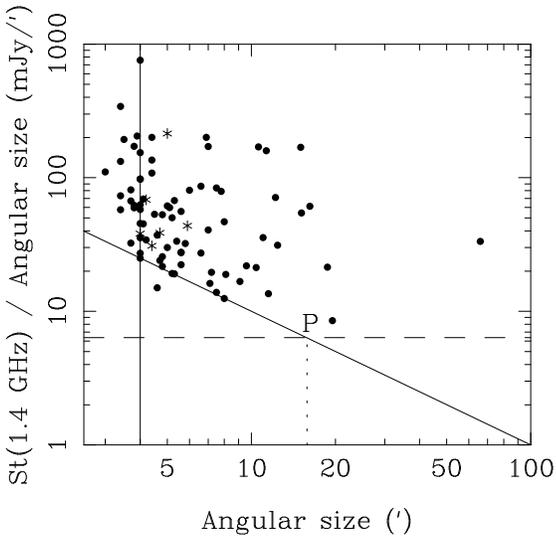}
%\rule{0.4pt}{4cm}% line thickness, height of picture
\caption{Integrated flux density per length unit against the source 
angular length. The vertical and the oblique solid lines represent the
sample limits in angular size ($\ge 4\arcmin$) and flux density
(S$_t^{1.4}\ge $100 mJy), respectively. The dashed horizontal line
represents the sensitivity limit of the NVSS (see text).
The vertical dotted line marks the angular size of $15.8\arcmin$,
above which sources with a flux density greater than 100 mJy could be
undetected in the NVSS (P point). Sources with uncertain redshift in 
Table~\ref{sample} are represented with asterisks.}
\label{fig_fluxsize}
\end{figure}

In Fig.~\ref{fig_fluxsize} we represent the flux density per length
unit against the source angular size for all the members of the sample. In
this plot we can clearly see the limitations imposed by the selection
criteria. The vertical solid line represents the lower limit in
angular size ($4\arcmin$). As mentioned in Section 3, there are 13
sources with sizes below that limit, which have been kept in the
sample. The oblique solid line represents the lower limit in total
flux density (100 mJy). There is one source (J0525+718) which lies
below that limit (Section 3). The horizontal dashed line marks the limit
imposed by the NVSS sensitivity ($1\sigma = 0.45$ mJy/beam), computed
considering a rectangular source with a fixed width of $3\arcmin$ (an
upper limit inferred from the NVSS maps), a variable length $l$ and an
uniform brightness distribution at $3\sigma$. The intersection of this
line with the flux density limit line (P point) defines the 
length ($l \sim 16\arcmin$)
above which sources with flux densities above 100 mJy could be missed 
in our sample. 

In order to investigate if such very extended sources could be more
properly studied using a lower frequency survey, we have cross-checked
our sample with a sample of 47 low redshift (z$\leq 0.4$) GRGs with
angular sizes larger than $5\arcmin$ selected from the WENSS
(Schoenmakers \cite{arnotesis}). Two radio galaxies, J1047+747 and
J1308+619, appear in this sample and are not present in our sample due
to their low total flux densities at 1.4 GHz, but not to their too
large size.  The rest of the sources in Schoenmakers sample with
declination $\ge +60^{\circ}$ are also in our sample. On the
other hand, we find 8 low redshift GRGs with angular sizes larger than
$5\arcmin$ that are missing from Schoenmakers sample, most possibly
due to an underestimation of the true source size induced by the low
resolution of the WENSS survey (Schoenmakers priv. comm.). These 8 sources 
are J0607+612, J0926+653, J1216+674,
J1844+653, J1853+800, J1918+742, J1951+706 and J2035+680. We have
also checked that all giant radio galaxies larger than $4\arcmin$ in
the compilation by Ishwara-Chandra \& Saikia (\cite{ishwara}) are in
our sample.  Therefore, we have enough confidence that the selection
from the NVSS is a good procedure (at least as good as others) to
define samples of extended radio sources. In fact, we find in our
sample 22 new GRGs (written in boldface in Table~\ref{sample}),
increasing to a total of 103 the number of known giants.

\begin{figure}
\vspace{7cm}
\includegraphics{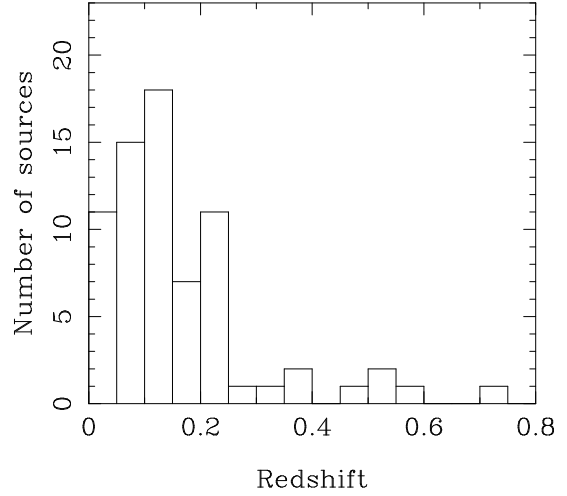}
%\rule{0.4pt}{4cm}% line thickness, height of picture
\caption{Histogram of the distribution of redshifts in our sample. The 
redshift bin is 0.05.} 
\label{fig_histoz}
\end{figure}

\begin{figure}
\vspace{7cm}
\includegraphics{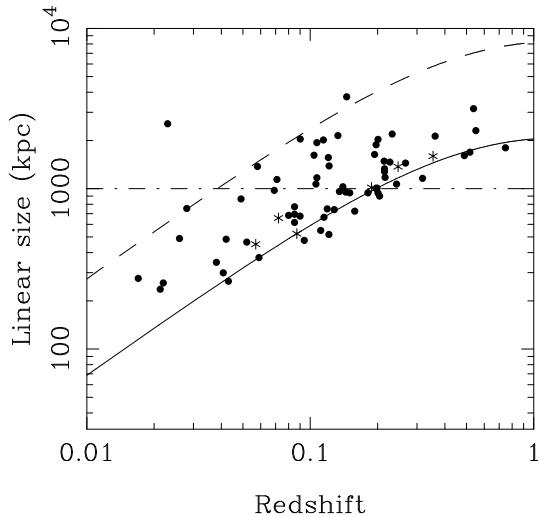}
%\rule{0.4pt}{4cm}% line thickness, height of picture
\caption{Projected linear distance against redshift. The solid line indicates the limit imposed by our selection criteria (as in Fig.~\ref{fig_fluxsize}, there are some sources smaller than 4$\arcmin$). The dashed line represents the sensitivity limit of the NVSS for a 16$\arcmin$ extended 100 mJy source (see text). The horizontal
 dotted line marks the usual definition of giant radio galaxies. Sources with uncertain redshift in 
Table~\ref{sample} are represented with asterisks.} 
\label{fig_dz}
\end{figure}

Fig.~\ref{fig_histoz} shows the number of sources per redshift bin of
0.05. We find that 87\% of the sources with known redshift are below
z$=0.25$. In fact, the selection criteria require sources with z$\ge 0.5$
to have projected linear sizes larger than 1.7 Mpc (see Fig.~\ref{fig_dz}).
Since such huge sources are rare (eg. Ishwara-Chandra \& Saikia
\cite{ishwara}), it is not unexpected that our sample is mostly composed by 
relatively nearby radio galaxies.

\begin{figure}
\vspace{7cm}
\includegraphics{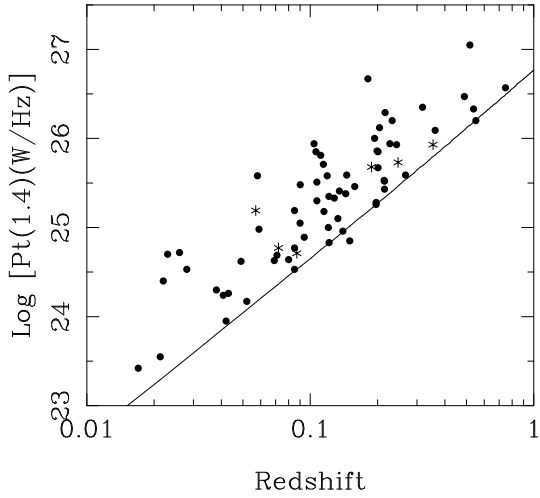}
%\rule{0.4pt}{4cm}% line thickness, height of picture
\caption{Luminosities of the members of the sample plotted against their redshifts. The solid line represents the flux density limit imposed by our selection criteria (S$_t^{1.4}\ge$100 mJy). Sources with uncertain redshift in 
Table~\ref{sample} are represented with asterisks.} 
\label{fig_pz}
\end{figure}

The total radio power at 1.4 GHz as a function of the source redshift
is represented in Fig.~\ref{fig_pz}.  The effect of the total flux density
limitation (S$_t^{1.4} \ge $100 mJy) is easily visible and shown by a
solid line.  This limitation masks any possible trend of the radio
power distribution with the redshift, although we note the small number of
nearby (z $\le 0.1$) high power ($\log$\,P$_{t}^{1.4} \ge 25.5$) radio
sources in the sample. Similarly, if we represent the source linear
sizes against their redshifts (Fig.~\ref{fig_dz}), we find a small
number of giant radio galaxies with z $\le 0.1$. We plot in 
Fig.~\ref{fig_dz} a dashed line which represents the locus
of a 100 mJy $16\arcmin$ large radio source in this diagram (P point
in Fig.~\ref{fig_fluxsize}), to give an idea of the sensitivity limit of
the NVSS. We find that GRGs with z$\le 0.04$ could be below the detection
limit. However, the dashed line does not define a stringent limit
since the real brightness distribution of radio sources is not
rectangular and uniform as we have previously assumed.  In
consequence, much larger sources could be
detected and, in fact, we find a low redshift source in our sample
above that limit: J1632+825 (NGC~6251; dominated by a strong narrow
jet).

The lack of small size sources at high redshift is entirely related to
our selection limit of 4$\arcmin$ in size (solid line; there are
sources below the line, which correspond to sources smaller than
4$\arcmin$ in Table~\ref{sample}). Moreover, we have a few giant radio
galaxies at high redshift (z$\ge 0.5$).  From previous studies
(eg. Schoenmakers et al. \cite{arno2}, Lara et al. \cite{lara3}) it
was derived that giant sources are characterized by large spectral
ages. The existence of giant radio galaxies at high redshift would
then imply that old radio sources are present at z $\ge 0.5$.

Another question raised after a first inspection of the sample is why
we do not find radio sources with linear sizes below 200 Kpc. From
Fig.~\ref{fig_dz} we find that intrinsically small sources should have
redshifts below 0.03 to fit the size limit of 4$\arcmin$. We have 6
sources below this redshift, but with angular sizes above
7$\arcmin$. The absence of small nearby sources must be related with
the radio galaxy population and the small volume enclosed at low
redshifts.  There exist indeed sources smaller than 100 Kpc which fit
our requirements, but must be rare (one example should be Cen A, if
it were at declinations above $+60^{\circ}$).

\begin{figure}
\vspace{7cm}
\includegraphics{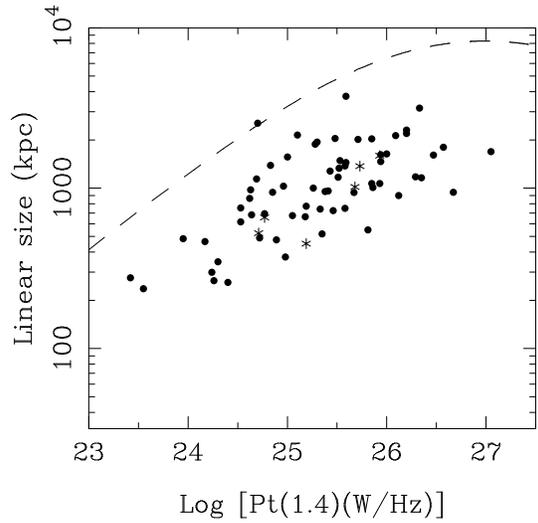}
%\rule{0.4pt}{4cm}% line thickness, height of picture
\caption{Projected linear sizes of the members of the sample plotted
against their luminosities. The dashed line represents the same sensitivity 
limit shown in Fig.~\ref{fig_dz}. Sources with uncertain redshift in 
Table~\ref{sample} are represented with asterisks.}
\label{fig_dp}
\end{figure}

In Fig.~\ref{fig_dp} we have plotted the source projected linear size versus 
the source total radio power at
1.4 GHz. The dotted line
represents the ``limit'' due to sensitivity limitations of the NVSS,
as in Fig.~\ref{fig_dz}. We find a striking absence of intrinsically
small high power radio galaxies. All sources with sizes below 500 Kpc
have $\log$P$_{t}$(1.4) below 25.5. Since our criteria require that sources
smaller than 500 Kpc be at z$\le 0.1$, this is not a selection
effect and must be a consequence of the statistically low number of
sources with high radio power enclosed in the limited volume given by
z $<$ 0.1. Nearby sources have, in general, low power.

\begin{figure}
\vspace{7cm}
\includegraphics{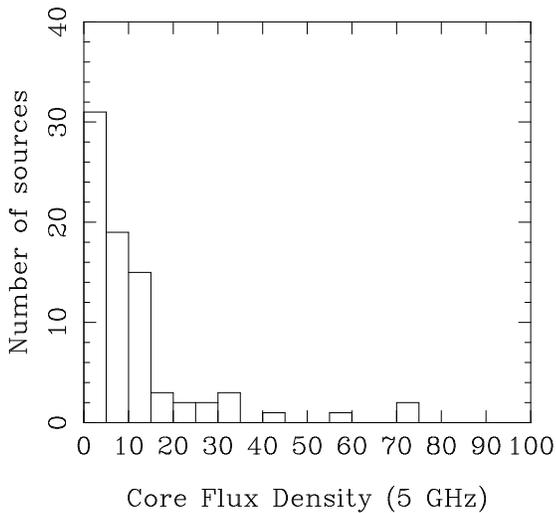}
%\rule{0.4pt}{4cm}% line thickness, height of picture
\caption{Histogram of the core flux density of the sample sources at 4.9 GHz, 
up to 100 mJy. The bin size is 5 mJy.} 
\label{fig_histosc}
\end{figure}

Finally, we plot in Fig.~\ref{fig_histosc} a histogram of the core
flux density at 4.9 GHz, up to 100 mJy. There are 5 sources with a
core flux density above this value and therefore not represented in
this plot. We note that our selection criteria do not impose
restrictions on the core flux density and in fact, 51 out of 84 radio
sources have weak cores, below 10 mJy.  This is an expected
result, since our sample selection criteria do not favor jets pointing
toward the observer and, consequently, we do not expect relativistic
Doppler boosting of intrinsically weak cores.  However, the study of
the parsec scale properties of the members of the sample by means of
VLBI observations will require, in most cases, the use of the phase
referencing technique in order to achieve enough sensitivity to
properly map radio cores with a flux density below 10-20 mJy.

\section{Conclusions}

We have presented in this paper a new sample of 84 large angular size
radio galaxies selected from the NVSS, with total flux density at 1.4
GHz $\ge 100$ mJy and declination above $+60^{\circ}$.  The selection
has been made using high resolution and sensitive VLA observations at
1.4 and 4.9 GHz, which allowed us to properly map extended structures
and to determine the radio core position. We measured new redshifts for 
44 radio sources (Paper II).

We have discussed selection effects and limits. Within our selection
criteria the final sample is homogeneous and can be used for
statistical studies (Paper III).

37 radio sources in our sample have a linear size larger than 1 Mpc,
of which 22 are new.  Adding to these the sample of GRGs by
Schoenmakers (\cite{arnotesis}) and the compilation of GRGs from
literature data by Ishwara-Chandra \& Saikia (\cite{ishwara}), the
number of known GRGs rises to 103, which allows a detailed study of
the properties of this poorly known class of radio sources.

\begin{acknowledgements}

This research is supported in part by the Spanish DGICYT
(PB97-1164). GG and LF acknowledge the Italian Ministry for University and
Research (MURST) for financial support under grant Cofin98-02-32. The
National Radio Astronomy Observatory is a facility of the National
Science Foundation operated under cooperative agreement by Associated
Universities, Inc.  This research has made use of the NASA/IPAC
Extragalactic Database (NED) which is operated by the Jet Propulsion
Laboratory, California Institute of Technology, under contract with
the National Aeronautics and Space Administration.

\end{acknowledgements}

\end{document}